\documentclass[english,aps,manuscript]{revtex4}
\usepackage[T1]{fontenc}
\usepackage[utf8]{inputenc}
\usepackage{graphicx}
\usepackage{esint}

\makeatletter
\@ifundefined{textcolor}{}
{%
 \definecolor{BLACK}{gray}{0}
 \definecolor{WHITE}{gray}{1}
 \definecolor{RED}{rgb}{1,0,0}
 \definecolor{GREEN}{rgb}{0,1,0}
 \definecolor{BLUE}{rgb}{0,0,1}
 \definecolor{CYAN}{cmyk}{1,0,0,0}
 \definecolor{MAGENTA}{cmyk}{0,1,0,0}
 \definecolor{YELLOW}{cmyk}{0,0,1,0}
}

\let\jnl@style=\rm
\def\ref@jnl#1{{\jnl@style#1}}

\def\aj{\ref@jnl{AJ}}                   
\def\actaa{\ref@jnl{Acta Astron.}}      
\def\araa{\ref@jnl{ARA\&A}}             
\def\apj{\ref@jnl{ApJ}}                 
\def\apjl{\ref@jnl{ApJ}}                
\def\apjs{\ref@jnl{ApJS}}               
\def\ao{\ref@jnl{Appl.~Opt.}}           
\def\apss{\ref@jnl{Ap\&SS}}             
\def\aap{\ref@jnl{A\&A}}                
\def\aapr{\ref@jnl{A\&A~Rev.}}          
\def\aaps{\ref@jnl{A\&AS}}              
\def\azh{\ref@jnl{AZh}}                 
\def\baas{\ref@jnl{BAAS}}               
\def\bac{\ref@jnl{Bull. astr. Inst. Czechosl.}}
\def\caa{\ref@jnl{Chinese Astron. Astrophys.}}
\def\cjaa{\ref@jnl{Chinese J. Astron. Astrophys.}}
\def\icarus{\ref@jnl{Icarus}}           
\def\jcap{\ref@jnl{J. Cosmology Astropart. Phys.}}
\def\jrasc{\ref@jnl{JRASC}}             
\def\memras{\ref@jnl{MmRAS}}            
\def\mnras{\ref@jnl{MNRAS}}             
\def\na{\ref@jnl{New A}}                
\def\nar{\ref@jnl{New A Rev.}}          
\def\pra{\ref@jnl{Phys.~Rev.~A}}        
\def\prb{\ref@jnl{Phys.~Rev.~B}}        
\def\prc{\ref@jnl{Phys.~Rev.~C}}        
\def\prd{\ref@jnl{Phys.~Rev.~D}}        
\def\pre{\ref@jnl{Phys.~Rev.~E}}        
\def\prl{\ref@jnl{Phys.~Rev.~Lett.}}    
\def\pasa{\ref@jnl{PASA}}               
\def\pasp{\ref@jnl{PASP}}               
\def\pasj{\ref@jnl{PASJ}}               
\def\rmxaa{\ref@jnl{Rev. Mexicana Astron. Astrofis.}}%
\def\qjras{\ref@jnl{QJRAS}}             
\def\skytel{\ref@jnl{S\&T}}             
\def\solphys{\ref@jnl{Sol.~Phys.}}      
\def\sovast{\ref@jnl{Soviet~Ast.}}      
\def\ssr{\ref@jnl{Space~Sci.~Rev.}}     
\def\zap{\ref@jnl{ZAp}}                 
\def\nat{\ref@jnl{Nature}}              
\def\iaucirc{\ref@jnl{IAU~Circ.}}       
\def\aplett{\ref@jnl{Astrophys.~Lett.}} 
\def\apspr{\ref@jnl{Astrophys.~Space~Phys.~Res.}}
\def\bain{\ref@jnl{Bull.~Astron.~Inst.~Netherlands}} 
\def\fcp{\ref@jnl{Fund.~Cosmic~Phys.}}  
\def\gca{\ref@jnl{Geochim.~Cosmochim.~Acta}}   
\def\grl{\ref@jnl{Geophys.~Res.~Lett.}} 
\def\jcp{\ref@jnl{J.~Chem.~Phys.}}      
\def\jgr{\ref@jnl{J.~Geophys.~Res.}}    
\def\jqsrt{\ref@jnl{J.~Quant.~Spec.~Radiat.~Transf.}}
\def\memsai{\ref@jnl{Mem.~Soc.~Astron.~Italiana}}
\def\nphysa{\ref@jnl{Nucl.~Phys.~A}}   
\def\physrep{\ref@jnl{Phys.~Rep.}}   
\def\physscr{\ref@jnl{Phys.~Scr}}   
\def\planss{\ref@jnl{Planet.~Space~Sci.}}   
\def\procspie{\ref@jnl{Proc.~SPIE}}   

\makeatother

\usepackage{babel}
\begin{document}

\title{Discrete Self Similarity in Filled Type I Strong Explosions}

\author{Almog Yalinewich$^{1}$ and Re'em Sari$^{1,2}$}

\address{$^{1}$Racah Institute of Physics, the Hebrew University, 91904,
Jerusalem, Israel\\
$^{2}$California Institute of Technology, MC 130-33, Pasadena, CA
91125}
\begin{abstract}
We present new solutions to the strong explosion problem in a non
power law density profi{}le. The unperturbed self similar solutions
developed by Sedov, Taylor and Von Neumann describe strong Newtonian
shocks propagating into a cold gas with a density profi{}le falling
off as $r^{-\omega}$, where $\omega\le\frac{7-\gamma}{\gamma+1}$
(filled type I solutions), and $\gamma$ is the adiabatic index of
the gas. The perturbations we consider are spherically symmetric and
log periodic with respect to the radius. While the unperturbed solutions
are continuously self similar, the log periodicity of the density
perturbations leads to a discrete self similarity of the perturbations,
i.e., the solution repeats itself up to a scaling at discrete time
intervals. We discuss these solutions and verify them against numerical
integrations of the time dependent hydrodynamic equations. This is
an extension of a previous investigation on type II solutions and
helps clarifying boundary conditions for perturbations to type I self
similar solutions.
\end{abstract}
\maketitle

\section{Introduction}

Expanding shock waves are naturally produced by diverse astrophysical
phenomena, such as supernovae, gamma ray bursts and stellar winds.
So far, analytical self similar solutions have been found for several
simple cases, of which we take special interest in the case of strong
spherical shocks propagating into a density profi{}le that decays
as a power of the radius
\begin{equation}
\rho_{a}\left(r\right)=kr^{-\omega}\label{eq:unpert_density}
\end{equation}
 The first solutions of this kind to be found, now commonly known
as the Sedov Taylor Von-Neumann solutions \cite{Taylor1950}, for
the case $\omega<3$ describe decelerating shocks. The solutions are
based on the conservation of energy inside the shocked region, and
they are called type I solutions. If $\omega<\frac{7-\gamma}{\gamma+1}$,
where $\gamma$ is the adiabatic index of the ambient gas, then the
explosion is filled, i.e. the pressure is greater than zero anywhere
inside the shocked region. If $\frac{7-\gamma}{\gamma+1}<\omega<3$,
then the explosion is hollow, i.e. the pressure (and the density)
vanish at a finite radius \cite{Waxman1993}. If $\omega=\frac{7-\gamma}{\gamma+1}$,
then the hydrodynamic equations admit a relatively simple solution
known as the Primakoff solution \cite{Sedov1977}. If $\omega>3$
the energy diverges at the center, so energy conservation no longer
applies and a different condition must be used \cite{Waxman1993}.
In this paper we will focus on filled type I explosions ($\omega\le\frac{7-\gamma}{\gamma+1}$).

The solutions discussed above, while useful, falls short when describing
shocks propagating into density profiles that deviate from a simple
power law decay. This might occur in a variety of astrophysical scenarios.
One example could be the propagation of an outward shock wave in a
stratified core collapse supernova progenitor \cite{Janka2012}. Another
example might be the interaction of a supernova shock wave with a
circumstellar bubble \cite{Chevalier1989}. Such bubbles form around
progenitors that emit strong stellar wind that pushes the circumstellar
wind away, so when the shock emerges from the progenitor, it first
interacts with a low density medium inside the bubble, and later with
the higher density medium outside. One example that we will dwell
on is the variation of the luminosity due to the interaction of a
supernova shock wave with a heterogeneous interstellar material.

From the reasons mentioned above, one could understand the need to
generalize as much as possible the external density profile for which
we can obtain analytic solutions, and this is what we attempt here.
This paper takes after a similar endeavor for type II solutions \cite{Oren2009}.

The idea of applying perturbation theory to the strong explosion problem
is not new, but so far it focused on stability analysis. Throughout
the years it has stirred up many controversies, most of which regarding
inner boundary conditions. The stability of type I explosions was
first studied by Bernstein and Book \cite{Bernstein1980}, but their
analysis was later refuted by Gaffet \cite{Gaffet1984,Gaffet1984a}.
Consequently, a new perturbation theory was proposed by Ryu and Vishniac
\cite{Ryu1987,Vishniac1989,Ryu1991}. However, Kushnir and Waxman
pointed out a possible error with the analysis of Ryu and Vishniac,
and proposed yet another boundary condition to the perturbation theory
\cite{Kushnir2005}. Numerical simulations \cite{MacLow1993} and
experiments with high power lasers \cite{Edens2010} are in general
agreement with the results of Ryu and Vishniac. The bone of contention
in these controversies is the inner boundary conditions, i.e. the
value of the hydrodynamic variables at the center. This paper will
attempt to shed light on the question of the correct boundary conditions.

The plan in this paper is as follows: In Sec. II we review the unperturbed
solutions and the boundary conditions at the front and at the center.
In Sec. III we develop the perturbation equations and boundary conditions.
We then discuss the solutions to these equations and compare them
to numerical results obtained from a full hydrodynamic simulation,
and finally we conclude in Sec. IV.

\section{The Unperturbed Solutions}

We proceed to give a quick review of the unperturbed solutions under
considerations \cite{Sedov1977}. The physical scenario is the deposition
of a large amount of energy from a point source at the center of a
spherically symmetric distribution of cold gas. It may be noted that
spherical symmetry was chosen for its relevance to most astrophysical
scenarios, but planar an cylindrical geometries may readily be treated
as well. The gas density follows a power law behavior (equation \ref{eq:unpert_density}).

\subsection{The Hydrodynamic Equations}

We begin with the Euler equation for an ideal fluid with adiabatic
index $\gamma$ in spherical symmetry
\begin{equation}
\frac{\partial\rho}{\partial t}+\frac{1}{r^{2}}\frac{\partial}{\partial r}\left(r^{2}\rho u\right)=0
\end{equation}
\begin{equation}
\rho\frac{\partial u}{\partial t}+u\frac{\partial u}{\partial r}+\frac{\partial}{\partial r}\left(\frac{\rho c^{2}}{\gamma}\right)=0
\end{equation}

\begin{equation}
\left(\frac{\partial}{\partial t}+u\frac{\partial}{\partial r}\right)\ln\left(\frac{c^{2}}{\gamma\rho^{\gamma-1}}\right)=0
\end{equation}
These equations feature the density $\rho$, velocity $u$ and speed
of sound $c$ as the dependent variables. They are usually expressed
in terms of the pressure $p$ rather than the speed of sound, and
they are related by
\begin{equation}
c^{2}=\gamma\frac{p}{\rho}
\end{equation}
We define dimensionless variables
\begin{equation}
r=R\left(t\right)\xi
\end{equation}
\begin{equation}
u\left(r,t\right)=\dot{R}\xi U\left(\xi\right)
\end{equation}
\begin{equation}
c\left(r,t\right)=\dot{R}\xi C\left(\xi\right)
\end{equation}
\begin{equation}
\rho\left(r,t\right)=kR^{-\omega}G\left(\xi\right)
\end{equation}
\begin{equation}
p\left(r,t\right)=kR^{-\omega}\dot{R}^{2}P\left(\xi\right)
\end{equation}
where $R\left(t\right)$ is the shock radius. It is assumed that the
shock radius has power law dependence on time
\begin{equation}
R\left(t\right)=A\left(t-t_{0}\right)^{\alpha}\label{eq:shock trajectory}
\end{equation}

\subsection{Boundary Conditions}

The boundary conditions at the front are determined by the Rankine
Hugoniot shock conditions \cite{Landau1959}. 
\begin{equation}
U\left(\xi=1\right)=\frac{2}{\gamma+1}
\end{equation}
\begin{equation}
C\left(\xi=1\right)=\frac{\sqrt{2\gamma\left(\gamma-1\right)}}{\gamma+1}
\end{equation}
\begin{equation}
G\left(\xi=1\right)=\frac{\gamma+1}{\gamma-1}
\end{equation}
\begin{equation}
P\left(\xi=1\right)=\frac{2}{\gamma+1}
\end{equation}

The power law index $\alpha$ is determined by the boundary conditions
at the center. In principle, the center of an explosion can either
be a source or a sink of energy. If the energy injection is power
law of the time, than it is possible to obtain self similar solutions
\cite{Ryu1991}. It was shown that energy injection always creates
a hollow explosion \cite{Ryu1991}, as if the extra energy was the
work exerted by an expanding spherical piston. The condition that
the energy is conserved is therefore equivalent to the condition that
the velocity vanishes at the center. 

The total energy contained in the explosion is given by

\[
E=4\pi\int_{0}^{R}\left(\frac{1}{2}\rho u^{2}+\frac{p}{\gamma-1}\right)r^{2}dr\propto kR^{3-\omega}\dot{R}^{2}
\]
and the right hand side is independent of time only if
\begin{equation}
\alpha=\frac{2}{5-\omega}
\end{equation}

\subsection{Thin Shell Model}

As $\gamma\rightarrow1$, the compression (i.e. ratio between the
shocked and unshocked matter) increases, and matter is concentrated
into a thinner shell, while the interior contains gas with a finite
pressure and negligible density \cite{Ryu1987}. The density in the
shell diverges, but the surface mass density remains finite
\begin{equation}
\sigma=\frac{\rho_{a}\left(R\right)R}{3-\omega}
\end{equation}
The density in the interior (behind the shell) vanishes. The pressure
inside the shell is obtained from Rankine Hugoniot equations
\begin{equation}
p_{f}=\rho_{a}\left(R\right)\dot{R}^{2}
\end{equation}
but the pressure in the interior is
\begin{equation}
p_{i}=\frac{1}{2}\rho_{a}\left(R\right)\dot{R}^{2}
\end{equation}
this expression can be obtained from the implicit solution for the
dimensionless pressure as a function of the dimensionless velocity
\cite{Landau1959}. The material velocity at the front is equal to
the velocity of the shock
\begin{equation}
u_{f}=\dot{R}
\end{equation}

\noindent Since the density vanishes at the center, one might confuse
it with a hollow explosion. However, in hollow explosions the pressure
vanishes at a finite radius, while in this case the pressure remains
finite throughout.

\noindent We now turn to the energy balance of such explosion. Energy
can be distributed as either thermal or kinetic, and can be either
inside the shell or behind it. The kinetic energy behind the shell
is negligible because there's no mass there, and the thermal energy
of the shell is negligible because its volume is very small. As $\gamma\rightarrow1$,
the kinetic energy of the shell remains finite, but the thermal energy
behind the shell diverges, because it is proportional to $\left(\gamma-1\right)^{-1}$.
Hence most of the energy is concentrated behind the shell as thermal
energy. We can also use this approximation to find the relation between
the energy and the trajectory of the shock front
\begin{equation}
E=\frac{4\pi}{3}R^{3}\frac{p_{i}}{\gamma-1}=\frac{4\pi}{6}R^{3}\dot{R}^{2}\frac{\rho_{a}\left(R\right)}{\gamma-1}
\end{equation}
Substituting equation \ref{eq:shock trajectory} yields

\begin{equation}
A=\left[\left(\frac{5-\omega}{2}\right)^{2}\frac{6\left(\gamma-1\right)}{4\pi}\frac{E}{K}\right]^{1/\left(5-\omega\right)}
\end{equation}

\noindent We will later use this model to obtain analytic results
for perturbations in a gas with $\gamma\rightarrow1$. A relevant
question in this context is whether outside perturbations manage to
cross the thin, dense shell and affect the inner region. On the one
hand, the width of the shell goes to zero, but on the other hand,
so does the speed of sound. From mass conservation and the Rankine
Hugoniot relations, the width of the shell is
\begin{equation}
\frac{\Delta R}{R}=\frac{\gamma-1}{\left(\gamma+1\right)\left(3-\omega\right)}
\end{equation}
while the speed of sound at the shock front goes as
\begin{equation}
c_{f}=\frac{\sqrt{2\gamma\left(\gamma-1\right)}}{\gamma+1}\dot{R}
\end{equation}

\noindent so the time it takes for information to cross the shell
scales as $\sqrt{\gamma-1}$, and is therefore much smaller than the
time it takes the explosion to double its size when $\gamma\rightarrow1$.

\subsection{Primakoff Solution}

As was mentioned earlier, when $\omega=\frac{7-\gamma}{\gamma+1}$
the hydrodynamic equations admit a simple analytic solution
\begin{equation}
U=\frac{2}{\gamma+1}
\end{equation}
\begin{equation}
C=\frac{\sqrt{2\gamma\left(\gamma-1\right)}}{\gamma+1}
\end{equation}
\begin{equation}
G=\frac{\gamma+1}{\gamma-1}\xi
\end{equation}
\begin{equation}
P=\frac{2}{\gamma+1}\xi^{3}
\end{equation}
we will later see that for this solution it is possible to obtain
analytic solutions for the perturbation equations.

\section{Discrete Self Similar Perturbations}

\subsection{The Perturbation Equations}

We now come to the case of a perturbed density profile. For the perturbation
equation to be tractable we aim at a self similar solution by carefully
choosing a perturbation whose characteristic wavelength scales like
the radius. Namely, we take the perturbed density profile to be
\begin{equation}
\rho_{a}\left(r\right)+\delta\rho_{a}\left(r\right)=kr^{-\omega}\left(1+\varepsilon\left(\frac{r}{r_{0}}\right)^{q}\right)\label{eq:ambient_density_pert}
\end{equation}
where $r_{0}$ has dimensions of length and bears only on the phase
of the perturbation, $q$ is the growth rate of the perturbation and
$\varepsilon$ is a small, real and dimensionless amplitude. We take
the real part of any hydrodynamic complex quantity to be the physically
significant element.

\noindent We define perturbed flow variables
\begin{equation}
u\left(r,t\right)+\delta u\left(r,t\right)=\dot{R}\xi\left[U\left(\xi\right)+f\left(t\right)\delta U\left(\xi\right)\right]
\end{equation}
\begin{equation}
\rho\left(r,t\right)+\delta\rho\left(r,t\right)=kR^{-\omega}\left[G\left(\xi\right)+f\left(t\right)\delta G\left(\xi\right)\right]
\end{equation}
\begin{equation}
p\left(r,t\right)+\delta p\left(r,t\right)=kR^{-\omega}\dot{R}^{2}\left[P\left(\xi\right)+f\left(t\right)\delta P\left(\xi\right)\right]
\end{equation}
\begin{equation}
R\left(t\right)+\delta R\left(t\right)=R\left(t\right)\left[1+f\left(t\right)\right]
\end{equation}
To allow separation of variables, the function $f\left(t\right)$
must satisfy
\begin{equation}
f\left(t\right)=\frac{\varepsilon}{d}\left(\frac{R}{r_{0}}\right)^{q}\Rightarrow\frac{\dot{f}R}{f\dot{R}}=q
\end{equation}
Where the parameter $d$ represents the amplification of each mode,
and is determined by boundary conditions, as explained in the next
subsection. If $q$ is imaginary, the real part of $f\left(t\right)$
is periodic, the solution is discretely self similar, i.e. it repeats
itself up to a scaling factor in intervals of $\frac{\Delta R}{R}=\exp\left(\frac{2\pi}{\left|q\right|}\right)-1$.
While the unperturbed solution and the perturbations in their complex
form are both self similar, the physical solution which is the real
part of their sum is not.

\noindent Plugging the perturbed hydrodynamic variables into the hydrodynamic
equations yields dimensionless ODEs for the perturbed variables \cite{Oren2009}.

\subsection{Boundary Conditions for the Perturbations}

The boundary conditions for the perturbed variables at the blast front
are derived in a similar way to \cite{Chevalier1990,Ryu1987} and
are identical to those appearing in  \cite{Oren2009} 
\begin{equation}
\delta G\left(\xi=1\right)=\frac{\gamma+1}{\gamma-1}\left(d-\omega\right)-G'\left(1\right)\label{eq:pert_rankine_hugoniot_density}
\end{equation}
\begin{equation}
\delta U\left(\xi=1\right)=\frac{2}{\gamma+1}q-U'\left(1\right)\label{eq:pert_rankine_hugoniot_velocity}
\end{equation}
\begin{equation}
\delta P\left(\xi=1\right)=\frac{2}{\gamma+1}\left[2\left(q+1\right)-\omega+d\right]-P'\left(1\right)\label{eq:pert_rankine_hugoniot_pressure}
\end{equation}
In analogy to the unperturbed solution, where the parameter $\alpha$
is determined by the inner boundary conditions or total conservation
of energy, the parameter $d$ is determined by the same considerations.
Integration of the self similar ODEs from the front to center with
the wrong value of $d$ would yield non zero velocity at the center,
so the energy flux does not vanish, and the total energy is not conserved.
We recall that the energy flux is given by $u\left(\frac{\gamma}{\gamma-1}p+\frac{1}{2}\rho u^{2}\right)$,
but since the unperturbed density and velocity vanish at the center
in filled type I explosions, the first order contribution to the flux
would be $\frac{\gamma}{\gamma-1}p\cdot\delta u$ . Hence, it is sufficient
to require that $\delta u$ would vanish at the center. Near the center,
the derivatives of the self similar variables reduce to
\begin{equation}
\frac{d}{d\xi}\left(\frac{\delta U}{U}\right)=\frac{-q\left(\delta P/P\right)-3\left(\delta U/U\right)}{\xi}+O\left(\xi^{0}\right)
\end{equation}
\begin{equation}
\frac{d}{d\xi}\left(\frac{\delta P}{P}\right)=0+O\left(\xi^{0}\right)
\end{equation}

\noindent Hence for generic values, the pressure perturbation would
be constant, and the velocity perturbation would diverge as $\xi^{-3}$.
Recalling that the power radiated from the center is $r^{2}p\delta u\propto\xi^{3}\delta U$,
we see that choosing the wrong boundary condition would mean energy
transfer through the center (periodic, if $q$ is imaginary). The
condition for preventing the divergence of the velocity perturbation
is

\begin{equation}
\frac{\delta U\left(\xi=0\right)}{U\left(\xi=0\right)}=-\frac{q}{3}\frac{\delta P\left(\xi=0\right)}{P\left(\xi=0\right)}\label{eq:pert_centre_bc}
\end{equation}

\noindent In case of Primakoff explosions, the pressure also vanishes
at the center, so they require a different treatment (the energy also
doesn't change, but the conditions at the center are different). A
more detailed discussion of perturbations to Primakoff explosions
is given in section \ref{sub:primakoff_pert}.

\noindent We note that condition \ref{eq:pert_centre_bc} is different
from both \cite{Ryu1987} and \cite{Kushnir2005}. The reason is that
they treated angular perturbations, where the total energy of every
perturbation always averages out to zero after summing over all angles,
so energy considerations do not apply. The method of Ryu and Vishniac,
$\delta P\left(\xi=0\right)=0$, keeps the tangential velocity from
diverging, so it is irrelevant for radial perturbations. Thus, we
can understand why there should be two separate conditions for radial
and angular perturbations. We also note that in similar problem, e.g.
perturbations to type II explosions, the same inner boundary conditions
are used both for radial \cite{Oren2009} and angular perturbations
\cite{Sari2012}.

\subsection{The Discrete Self Similar Solution}

While self similarity simplifies the problem by reducing the PDEs
to ODEs, the resulting ODEs, in general, do not admit analytic solutions.
Therefore, for each specific set of parameters $\gamma$, $\omega$
and $q$, the functions $\delta G$, $\delta U$, $\delta P$ and
the parameter $d$ are found numerically. Since the ODEs are linear,
there exists a matrix that relates the vector of the values of the
flow variables at the center to the same vector at the front
\begin{equation}
\left(\begin{array}{c}
\delta G\left(1\right)\\
\delta P\left(1\right)\\
\delta U\left(1\right)
\end{array}\right)=\mathbf{M}\left(\begin{array}{c}
\delta G\left(0\right)\\
\delta P\left(0\right)\\
\delta U\left(0\right)
\end{array}\right)\label{eq:matix_equation}
\end{equation}
 It is possible to find this matrix numerically, since it is independent
of $d$. Thus equation \ref{eq:matix_equation} and the boundary conditions
constitutes 4 linear equation for 4 variables ($d$, $\delta G\left(0\right)$,
$\delta P\left(0\right)$ and $\delta U\left(0\right)$). Solving
these equations yields the value of $d$.

\begin{figure}
\begin{centering}
\includegraphics[width=9cm,height=9cm,keepaspectratio]{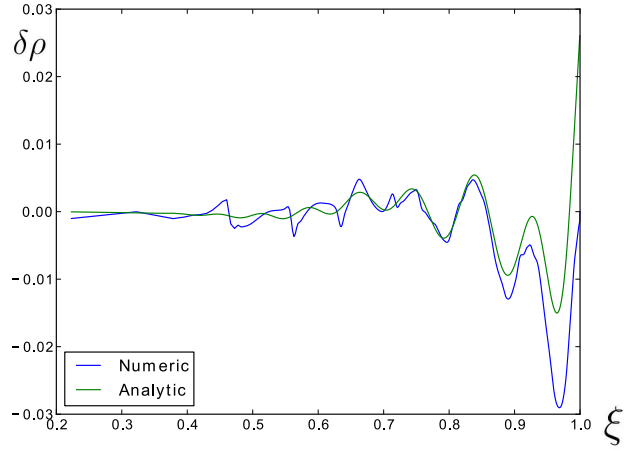}
\par\end{centering}

\begin{centering}
\includegraphics[width=9cm,height=9cm,keepaspectratio]{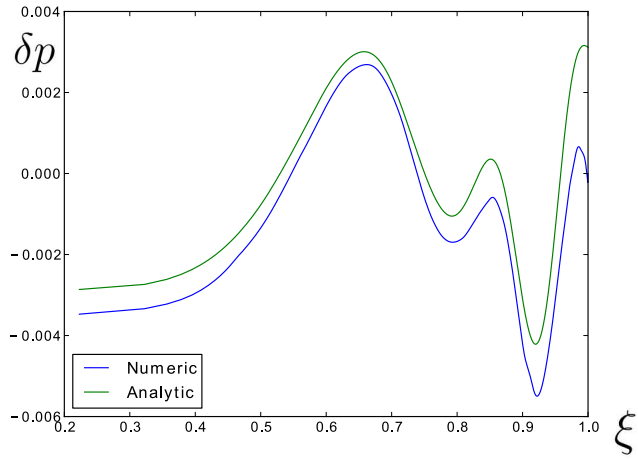}
\par\end{centering}

\begin{centering}
\includegraphics[width=9cm,height=9cm,keepaspectratio]{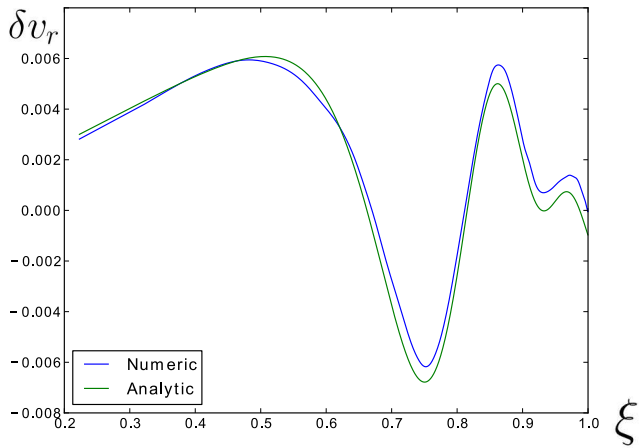}
\par\end{centering}

\caption{Comparison of the analytic and numeric profiles of the perturbed hydrodynamic
variables: density (top), pressure (middle) and velocity (bottom).
The explosion parameters are $\gamma=\frac{5}{3}$, $\omega=0$, $q=20i$,
$\varepsilon=0.01$ \label{fig:numeric_vs_analytic}}
\end{figure}

A comparison between the the solutions discussed above and a hydrodynamic
simulation is presented in figure \ref{fig:numeric_vs_analytic}.
All curves seem to agree. The numerical calculations were carried
out using the hydrocode PLUTO \cite{Mignone2007}. We have also verified
that better accuracy can be achieved by increasing the resolution.
However, infinite resolution will not reduce the error to zero, because
of differences between the initial conditions in the simulation and
those assumed in the mathematical formulation. One difference is the
size of the initial hot spot. In the mathematical problem the hot
spot is point like, while in the simulation it always has a finite
size. Another difference is the ambient pressure, which is assumed
to be zero in the mathematical problem, while in the simulation it
is also finite in the simulation.

Figure \ref{fig:numeric_vs_analytic} shows that the wavelength of
the density fluctuations is shorter than those of the pressure and
velocity. This happens because the density is affected by both traveling
sound waves and entropy waves, while the pressure and velocity are
affected solely by sound waves. From this argument it follows that
the characteristic wavelength are given by $\frac{2\pi}{\left|q\right|}\left(1-\xi U\pm\sqrt{\gamma\frac{P}{G}}\right)$
for the pressure and velocity, together with $\frac{2\pi}{\left|q\right|}\left(1-\xi U\right)$
for density perturbations.

\begin{figure}
\begin{centering}
\includegraphics[width=9cm,height=9cm,keepaspectratio]{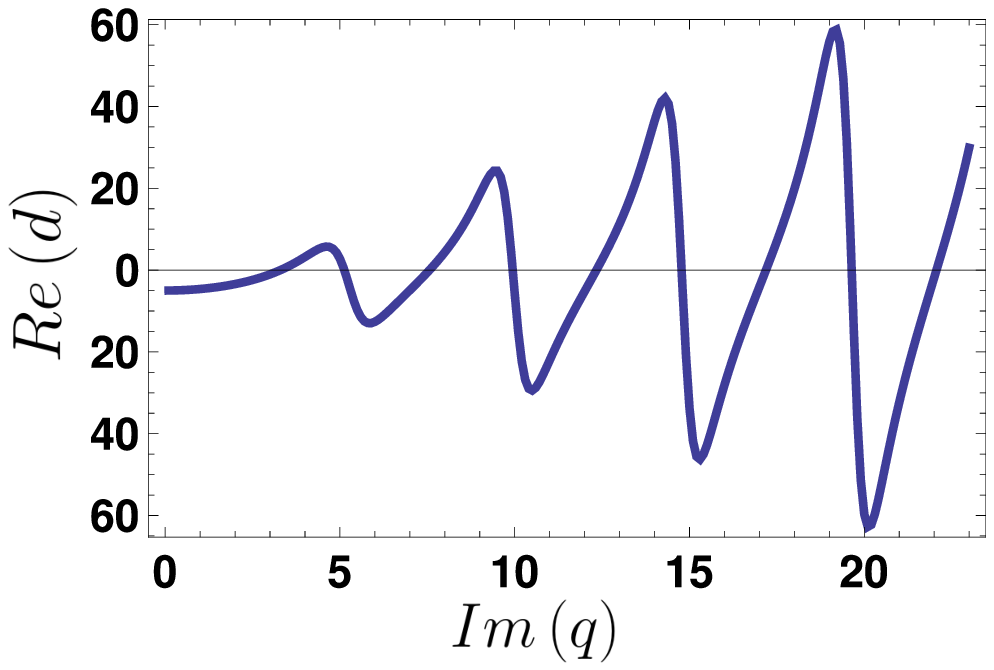}
\par\end{centering}

\begin{centering}
\includegraphics[width=9cm,height=9cm,keepaspectratio]{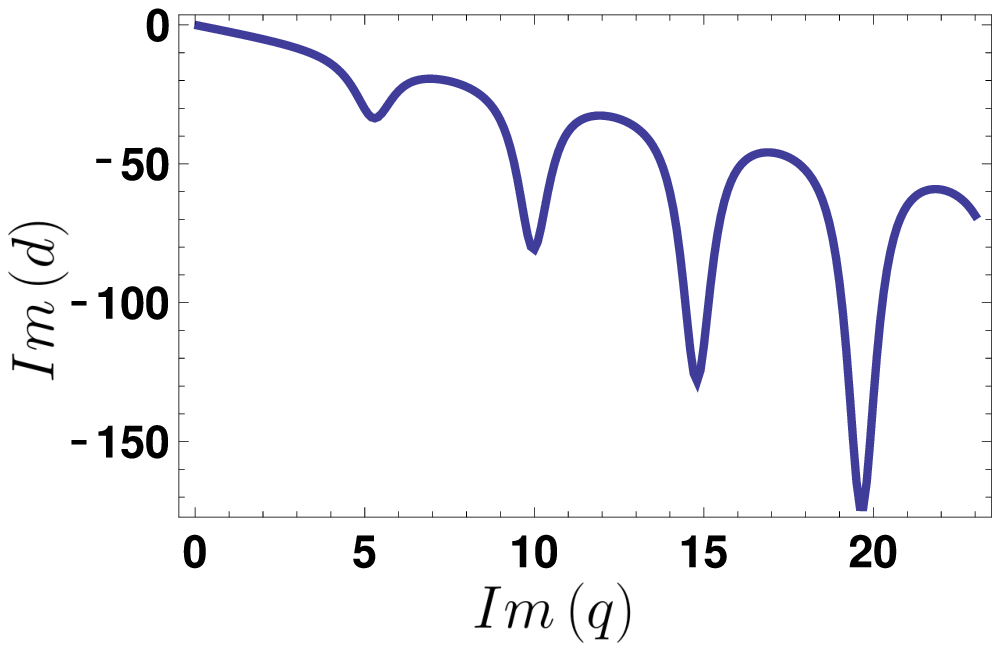}
\par\end{centering}

\caption{The real (top) and imaginary (bottom) part of $d$ as a function of
$Im\left(q\right)$, for an explosion with $\gamma=\frac{5}{3}$ and
$\omega=0$\label{fig:d_vs_b_filled}}

\end{figure}

\begin{figure}
\begin{centering}
\includegraphics[width=9cm,height=9cm,keepaspectratio]{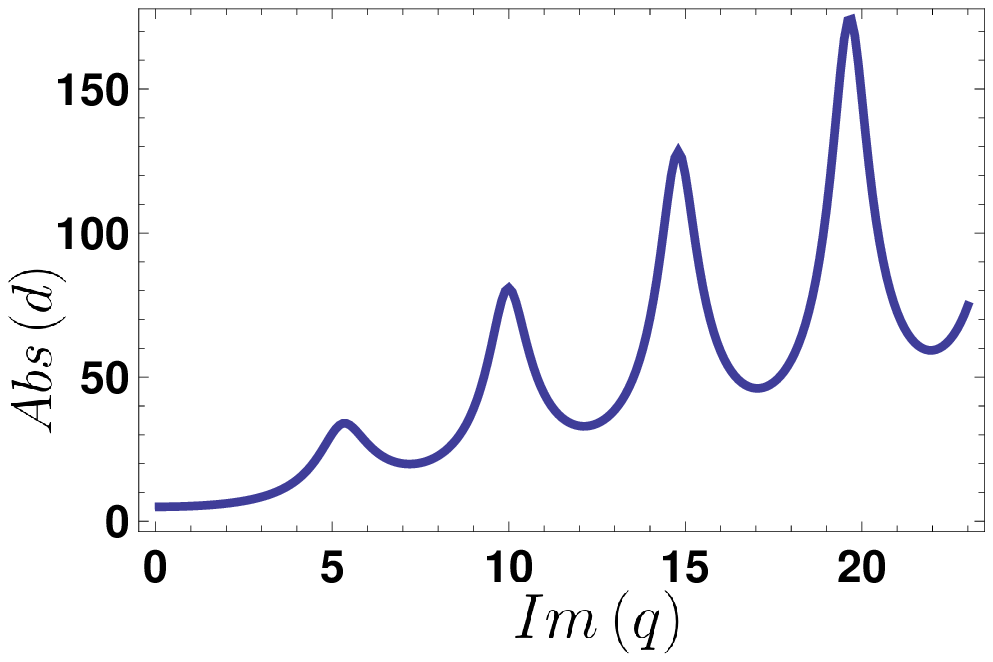}
\par\end{centering}

\begin{centering}
\includegraphics[width=9cm,height=9cm,keepaspectratio]{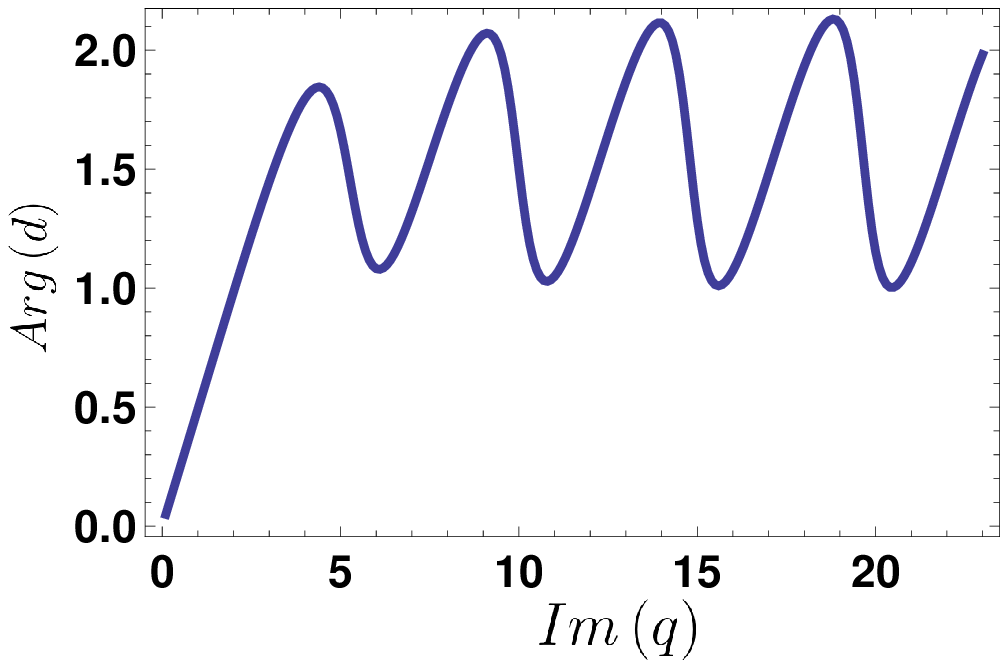}
\par\end{centering}

\caption{Absolute value (top) and phase (bottom) of the parameter $d$ as a
function of $Im\left(q\right)$ for an explosion with $\gamma=5/3$
and $\omega=0$.\label{fig:abs_arg_d_vs_b}}

\end{figure}

Finally, figures \ref{fig:d_vs_b_filled} and \ref{fig:abs_arg_d_vs_b}
show $d$ as a function of $Im\left(q\right)$, relating the fractional
perturbation in the shock position to the fractional perturbation
in the external density, for $\omega=0$ and $\gamma=\frac{5}{3}$.
The oscillations are due to the diffraction of the incident wave from
the blast front, with wave reflected from the center. This property
is qualitatively different from the behavior of the same curves plotted
for type II explosions \cite{Oren2009}. In type II explosions, sound
waves mostly travel from the front to sonic point, and not the other
way around, and that is why the $d\left(Im\left(q\right)\right)$
curves for type II explosions are monotonous. 

\noindent 

\noindent 

\noindent 

\subsection{Long Wavelength Limit}

Perturbations with $q=0$ correspond to perturbations in the coefficient
$K$ of the ambient density (equation \ref{eq:unpert_density}). From
units considerations we know that $E\propto KA^{5-\omega}$, so if
the energy is conserved $A\propto K^{\frac{1}{\omega-5}}$ and
\begin{equation}
d\left(q=0\right)=\omega-5\label{eq:d_q0}
\end{equation}

\subsection{Thin Shell Model}

In the thin shell model ($\gamma\rightarrow1$) the total energy is
given by $\frac{4\pi}{3}R^{3}p_{i}=\frac{4\pi}{6}R^{3}\rho_{a}\left(R\right)\dot{R}^{2}$.
From the conditions that the energy remains constant $\delta\left(R^{3}\rho_{a}\dot{R}^{2}\right)=0$,
we obtain the relation
\begin{equation}
d=\omega-5-2q\label{eq:d_thin_shell}
\end{equation}
In the limit $q\rightarrow0$ equation \ref{eq:d_thin_shell} reduces
to \ref{eq:d_q0}.

\subsection{Primakoff Solution\label{sub:primakoff_pert}}

In the case of the Primakoff explosion, the perturbation equations
can be solved analytically. With the substitution
\begin{equation}
\mathbf{Y}=\left(\frac{\delta G}{G},\frac{\delta P}{P},\frac{\delta U}{U}\right)^{T}
\end{equation}
the system of ODEs can be reduced to the form
\begin{equation}
\frac{d\mathbf{Y}}{d\ln\xi}=\mathbf{M}\cdot\mathbf{Y}
\end{equation}
\begin{equation}
\mathbf{M}=\left(\begin{array}{ccc}
\frac{6\left(\gamma-1\right)+q\left(\gamma+1\right)^{2}}{\gamma^{2}-1} & -\frac{2\left(-3+q+3\gamma+q\gamma\right)}{\gamma^{2}-1} & -\frac{2\left(7+q-\gamma+q\gamma\right)}{\gamma^{2}-1}\\
\frac{6\gamma}{\gamma+1} & -\frac{q+6\gamma+q\gamma}{\gamma+1} & -\frac{2\left(-3+\left(q+5\right)\gamma+q\gamma^{2}\right)}{\gamma^{2}-1}\\
\frac{3\left(\gamma-1\right)}{\gamma+1} & -\frac{-3+q+3\gamma+q\gamma}{\gamma+1} & -\frac{11+q+3\gamma+q\gamma}{\gamma+1}
\end{array}\right)
\end{equation}
The solution is
\begin{equation}
\mathbf{Y}\left(\xi\right)=\exp\left(\mathbf{M}\ln\xi\right)\mathbf{Y}\left(1\right)
\end{equation}
Every term in $\mathbf{Y}\left(\xi\right)$ is the sum of 3 power
laws in $\xi$, where each power is an eigenvalue of $\mathbf{M}$.

\begin{figure}
\begin{centering}
\includegraphics[width=9cm,height=9cm,keepaspectratio]{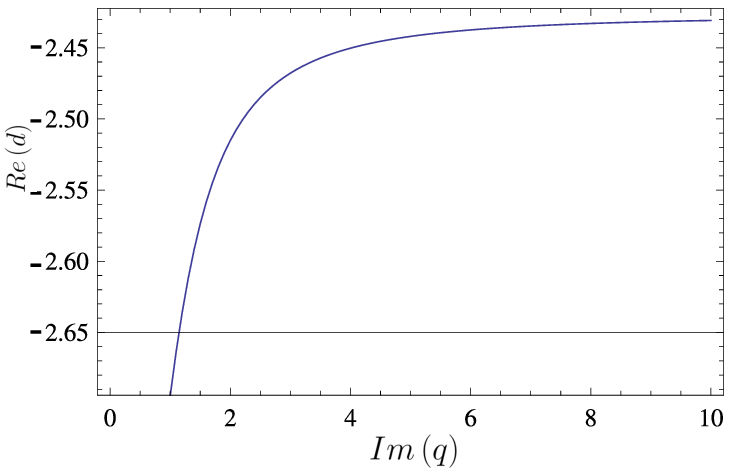}
\par\end{centering}

\begin{centering}
\includegraphics[width=9cm,height=9cm,keepaspectratio]{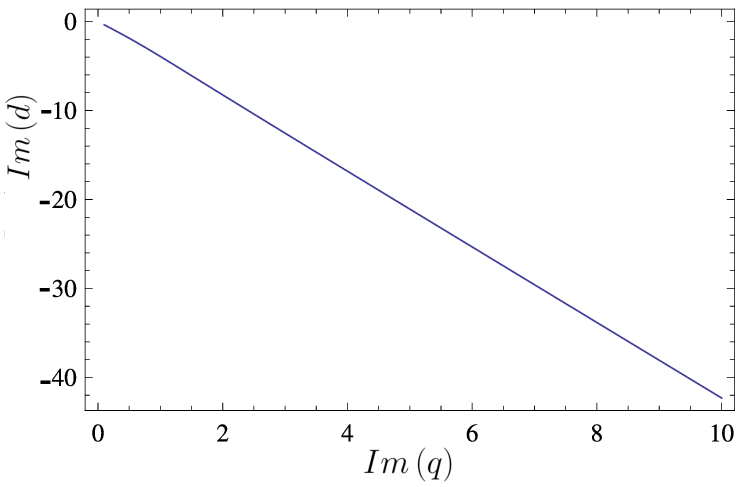}
\par\end{centering}

\caption{The real (top) and imaginary (bottom) parts of $d$, as a function
of $Im\left(q\right)$ , for $\gamma=\frac{5}{3}$ and $\omega=2$
(the Primakoff solution) \label{fig:prim_d_vs_q}}
\end{figure}

It is possible to perform the total energy integral explicitly for
this case. The parameter $d$ is chosen such that the total energy
remains the same. Another way to find $d$ by calculating the energy
flux at the center and requiring that it be equal to zero. Both ways
are mathematically equivalent, but the latter is computationally easier.
We were not able to obtain an explicit expression for the parameter
$d$, but for numerical values of $\gamma$, $\omega$ and $q$ the
parameter $d$ can be readily computed. The parameter $d$ as a function
of $Im\left(q\right)$ for $\gamma=\frac{5}{3}$ ($\omega=2$) is
given in figure \ref{fig:prim_d_vs_q}. We remark that that these
curves are monotonous, whereas we saw earlier that for smaller $\omega$
the graphs are oscillating. The reason is that there is no reflection
from the center in the case of Primakoff explosions, because the speed
of sound vanishes there. Therefore, the short wavelength limit discussed
in \cite{Oren2009} also applies to the Primakoff solution, so
\begin{equation}
\lim_{q\rightarrow\infty}\frac{d}{q}=-\sqrt{2+\frac{2\gamma}{\gamma-1}}
\end{equation}
The derivation of this result is based on the assumption that there
are no waves emanating from the center, so the outward going Riemann
invariant does not change. The same argument cannot be applied to
general filled type I explosions, because of the reflection from the
center.

\noindent 

\section{Discussion}

We have laid out a method for solving the strong explosion problem
in density profiles that deviate from a pure power law radial dependence.
The key lies in choosing radially log periodic perturbations which
do not introduce a new scale into the problem. This leads to self
similar perturbation in the hydrodynamic quantities behind the shock,
which can be found by solving a set of ordinary differential equations.
It is possible to obtain self similar equations for the perturbations
when the density perturbation is given in equation \ref{eq:ambient_density_pert},
but if $q$ is imaginary, then the solution is only discretely self
similar because of the periodic nature of the perturbations. We find
that the coefficient $d$ that relates the amplitude of the perturbations
in the shock position with the amplitude of the density perturbations
has a $O\left(1\right)$ real part and an $O\left(Im\left(q\right)\right)$
imaginary part, so at the short wavelength limit, $Im\left(q\right)\gg1$,
$\left|d\right|$ increases. From the boundary conditions at the shock
front (equations \ref{eq:pert_rankine_hugoniot_density}, \ref{eq:pert_rankine_hugoniot_velocity}
and \ref{eq:pert_rankine_hugoniot_pressure}) we see that the absolute
value of the dimensionless variables increases with $q$. The dimensional
perturbed variables are proportional to the dimensionless variables
divided by $d$, so at high values of $q$ their amplitudes tend to
a plateau.

The linearized perturbation treatment naturally ensures that the perturbations
will be linear in $\varepsilon$. This simplifies the solution of
the problem but limits the validity of the method to small perturbations.
The perturbation theory developed above fails when $\varepsilon$
becomes too large. The deviation from linear theory is of order $\varepsilon^{2}$.
It is possible to obtain a more quantitative assessment of the difference
by considering the long wavelength limit.

Since these perturbations are linear, it is possible to represent
arbitrary small deviations of a density profile from a power law by
a sum of different mode, as was done for type II solutions \cite{Oren2009}.

The crux of the problem discussed is choosing the correct inner boundary
conditions. The boundary conditions used here is different from both
that of Ryu \& Vishniac, and that of Kushnir \& Waxman. However, they
discussed angular perturbations, while we discuss radial perturbations
only, and we claim that the inner boundary conditions for radial perturbations
must be different from those of angular perturbations. The reason
is that radial inner boundary conditions are based on energy conservation,
which is irrelevant in angular perturbation as all modes conserve
energy.

We conclude with an example of an astrophysical relation: the relation
of a supernova remnant bolometric luminosity to density modulation
in the interstellar medium. Let us consider a supernova remnant shockwave
that propagates into the interstellar medium with a density $\rho_{a}$
distributed in the form of equation \ref{eq:ambient_density_pert}.
If the emitted flux would be some small fraction of the hydrodynamic
energy flux $\rho v^{3}$, the variation of the luminosity would be
\begin{equation}
\delta\ln L=\frac{\delta L}{L}=\delta\ln\left(\rho v^{3}R^{2}\right)=\frac{\delta\rho}{\rho}-3\frac{\delta v}{v}+2\frac{\delta R}{R}
\end{equation}

\noindent We give explicit results for the case $\omega=0$, $\gamma=5/3$
and use the approximation for a thin shockwave $d=\omega-5-2q$. From
equations \ref{eq:pert_rankine_hugoniot_density}, \ref{eq:pert_rankine_hugoniot_velocity}
and \ref{eq:pert_rankine_hugoniot_pressure} we get
\begin{equation}
\frac{\delta L}{L}=\frac{12+5q}{5+2q}\frac{\delta\rho_{a}}{\rho_{a}}
\end{equation}
This equation relates variations in the surrounding density to observed
flux. In the limit $q\rightarrow0$, where the wavelength of the perturbation
is long, the relative variation in the luminosity are 2.4 times larger
than the relative density variations, and both are in phase.

\bibliographystyle{plain}

\end{document}